\documentclass[10pt,journal,compsoc]{IEEEtran}
\usepackage{multirow}
\usepackage{hyperref}

\usepackage{mathptmx}
\usepackage{graphicx}
\usepackage{balance}  
\usepackage{graphics} 
\usepackage{times}    
\usepackage{url}      
\usepackage{amsmath}
\usepackage{mathtools}
\usepackage{cite}
\usepackage{csquotes}
\usepackage{array}
\usepackage{booktabs}
\usepackage{colortbl}
\usepackage{balance}  
\usepackage{booktabs}  
\usepackage{eqparbox}  
\usepackage{censor}
\usepackage{enumitem}

\usepackage{array}
\usepackage{fixltx2e}
\begin{document}
%

\title{HyFinBall: a Hybrid User Interface for Coordinated 2D+3D Visualization in Semi-Immersive VR}

%
%
%
%

\author{Isaac Cho, and~Zachary~Wartell
\IEEEcompsocitemizethanks{\IEEEcompsocthanksitem Isaac Cho is with the Department
of Computer Science, Utah State University, UT, 84319 \protect\\
\IEEEcompsocthanksitem Z. Wartell is with the Department of Computer Science, University of North Carolina at Charlotte, NC, 28223.}}

%
%

\markboth{Journal of \LaTeX\ Class Files,~Vol.~13, No.~9, September~2014}%
{Shell \MakeLowercase{\textit{et al.}}: Bare Demo of IEEEtran.cls for Computer Society Journals}
%



\IEEEtitleabstractindextext{%

\begin{abstract}
Sophisticated 3D visualization applications usually provide coordinated 2D and 3D views. Normally 3D input device is used for 3D tasks since they perform better than traditional 2D input devices. However, they do not perform better for 2D tasks. This paper presents a bimanual hybrid user interface that supports four interaction modes: a dual 6-degree-of-freedom (DOF) input device mode, a dual planar constrained 3DOF input device mode, a dual 2-finger multi-touch mode, and 3D hand and finger gestures. The application is a multi-dimensional visualization with coordinated 3D and 2D views on a desktop VR system. The input devices are buttonballs with seamless switching between 3D and 2D device modes, as well as between free-hand finger input and device usage. The 3D and 2D device mode switch automatically switches a buttonball's visual representation between a 3D cursor and a 2D cursor while changing the available user interaction techniques between 3D and 2D interaction techniques to interact with the coordinated views. The paper also provides two formal user studies to evaluate HyFinBall for various dimensional tasks, including 3D, 2D, and cross-dimensional tasks. Our experimental results show the benefits of the HyFinBall interface for cross-dimensional tasks that require 3D and 2D interactions.
\end{abstract}

}

\maketitle




\IEEEdisplaynontitleabstractindextext

%
\IEEEpeerreviewmaketitle

\IEEEraisesectionheading{\section{Introduction}\label{sec:introduction}}

\begin{figure*}
   \centering
   \includegraphics[width=0.9\textwidth]{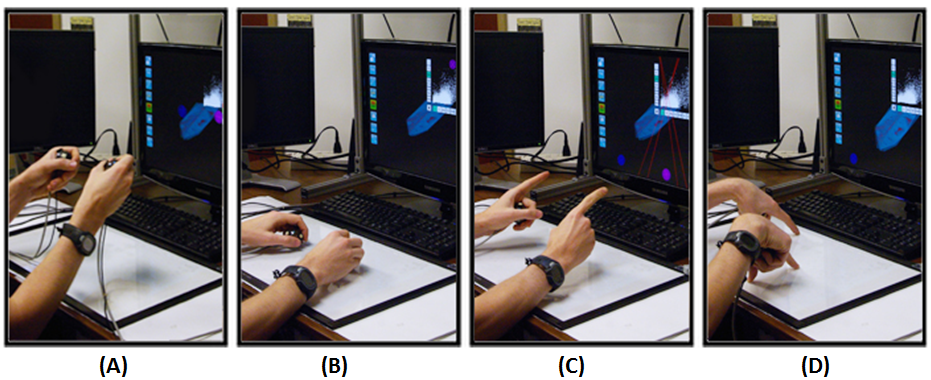}
   \caption{HyFinBall supports 6DOF isotonic input (A), planar-3DOF input (B), 3D hand and finger tracking and gesture (C), and multi-touch (D). Note, in these photos, the horizontal display is disabled to emphasize hand posture. In a running application, it would display 2D graphics (see Figure 2).}
    \label{fig:main}
\end{figure*}

\IEEEPARstart{S}{patial}  user interfaces (UIs) have been researched in various fields for many decades since Xerox Parc introduced the Windows-Icon-Menu-Pointer (WIMP) metaphor in the early 1970s. While mouse UI techniques play an important role in various visualizations and applications, the field still lacks a widely accepted standard for 3D UIs in 3D visualization. Researchers have developed numerous 3D interaction techniques to provide a higher degree-of-freedom (DOF) spatial input for 3D visualization and related applications ~\cite{laviola20173d, Bowman2008, Bowman2004}.

Sophisticated 3D visualization applications commonly integrate 3D and 2D visualizations as coordinated views. It is known that 3D input devices outperform traditional 2D input devices for 3D tasks in a 3D view. But it is also known that 2D input devices perform better than 3D input devices for 2D tasks. 

This paper introduces a hybrid user interface (HSU), HyFinBall, that supports various 2D and 3D interaction techniques for coordinated 2D+3D visualization. 
The HSU system is designed for a dual-screen Fish-tank VR setup \cite{Deering1992,Ware1997} to provide maximal physical support of the user's hands and arms. The HSU is designed to explore the following ideal:

\begin{displayquote}
\textit{A single interface (hardware and software) that supports both the best-of-breed input technology for 3D interaction and the best-of-breed input technology for 2D in a \textbf{seamless, flowing} manner that does \textbf{not} require dropping-and-picking up multiple devices to transition between 2D and 3D spatial input.}
\end{displayquote}

HyFinBall supports four modes (see Figure 1A through D): 

\begin{itemize}[noitemsep]
\item{(A) dual 6DOF input devices (xyz + yaw,pitch,roll)} 
\item{(B) dual planar constrained 3DOF input devices (xy + yaw)} 
\item{(C) 3D hand and finger gestures} 
\item{(D) dual two-finger multi-touch}
\end{itemize}

The 6DOF buttonballs and 3D finger-tracking enable 3D UI tasks, while the mouse-like 3DOF devices and multi-touch support 2D UI tasks. Each hand operates independently, allowing asymmetric or coordinated use. The HyFinBall hardware and software interface enables users to transition smoothly across all four input technologies without dropping or picking up different physical devices when shifting between spatial UIs. (Keyboard text entry, however, still requires setting down the buttonballs or using voice input.)

This paper contributes two controlled experiments that evaluate the hybrid user-interface capabilities of the HyFinBall system. Experiment 1 examines the usability of HyFinBall’s automatic switching between 6DOF input and planar-constrained 3DOF input when performing 3D and 2D tasks within a coordinated 2D–3D visualization environment. It compares this auto-switching condition with two additional device configurations: (1) a 6DOF-only condition, in which all tasks—both 2D and 3D—are performed using 6DOF input, and (2) a 6DOF+mouse condition, in which 3D and 2D interaction techniques are distributed across separate devices. All 12 participants successfully used HyFinBall’s auto-switch mechanism. Task completion times were lowest in the HyFinBall condition, and 10 of the 12 participants rated HyFinBall as the preferred method for performing the combined 2D–3D tasks.

Experiment 2 examines the usability of cycling between multi-touch tasks on a horizontal display surface (Figure 1D) and 3D tasks with the buttonball on a vertical display (Figure 1A) while ‘palming’ the buttonball—that is, keeping it tucked in the user’s palm during multi-touch interaction. This technique eliminates device-acquisition time penalties that would otherwise arise from repeatedly dropping and picking up the buttonball, and would not be feasible with bulkier 3D devices such as joystick-style controllers. All 16 participants successfully operated in the palming condition, and the results show reduced overall task completion times and faster response times for both 3D and multi-touch tasks during cross-dimensional interaction.


\section{Background and prior work}

3D UI techniques have a long history. Bowman et al. \cite{Bowman2004, Bowman2008} present a broad and detailed discussion on 3D UIs and interaction techniques. They identify specifications and heuristics of interaction techniques that can improve the usability of 3D interactions in real-world applications and present guidelines for developing new interaction techniques. Liu et al. explore modern interaction techniques for 3D desktop personal computers (PCs) \cite{Liu2000}. A number of other articles also include a review of physical input devices for 3D UIs \cite{Hinckley2002, Kruijff2006} and interaction techniques for large displays \cite{Grossman2001}.

Bimanual interaction enriches interaction because humans often use two hands to accomplish tasks in the real world. A significant amount of research shows the advantages of bimanual interactions \cite{Balakrishnan2000,Balakrishnan1999, Buxton1986} based on Guiard's Kinetic Chain theory that classifies different categories of bimanual actions \cite{Guiard1987}.

Several taxonomies of spatial input technologies (hardware) \cite{Buxton1986} have been created, as well as taxonomies of 3D spatial user interaction techniques (software) \cite{Mackinlay1990}. A 2D input device only tracks within a physical plane. 3D input tracks motion in 3-dimensions (at least 3DOF position typically up to 6DOF). Held devices are spatial input devices held by the user, while body-tracking tracks the body (such as hands and fingers). Body-tracking never requires the user to grasp a prop, but it may or may not require some encumbering mechanism (gloves, fiducial markers, etc.).

A traditional mouse is a 2D held device with 2 position DOFs. We denote a 2D mouse with the ability to yaw perpendicular to the motion plane \cite{Mackinlay1990} as a \enquote{planar-3DOF} device. Multi-touch is a body-tracking, 2D input with roughly 20 DOFs (10 fingers $\times$ 2 position DOFs). VIDEOPLACE was an early body-tracked 2D interface \cite{Krueger1985}. Notably, the VIDEOPLACE user was completely unencumbered (i.e., requiring no worn apparatus of any kind, not even fiducial markers). 

3D input interacts in a 3D space. The bat \cite{Ware1988} is an isotonic, 3D held-device with 6DOF pose (position and orientation). A bending-sensing data glove with a 6DOF tracker attached is categorized as 3D body-tracking, not a held device. The ideal implementation of body-tracking, of course, is a completely unencumbered system. Wang et al. \cite{Wang2011} demonstrate unencumbered hand and finger tracking, and commercial products such as the Kinect, LeapMotion, and 3Gear are now available.

Various researchers in 3D UI and tangible interfaces \cite{Ishii1997} have demonstrated that having a held device grasped in the hand is beneficial due to the tactile feedback (passive haptics) it provides for 2D and 3D manipulation. Such feedback does not exist in hand or finger-tracked 3D UIs, but does exist in 2D multi-touch UI's or 3D systems augmented with active (robotic) haptics.

When considering a held input device, a device is held in either a precision grasp or a power grasp. For some applications, such as a VR system for training a user to use a real-world tool, a power-grasp prop would be ideal, assuming the real-world tool requires a power-grasp. However, a precision grip allows finer control due to the larger \enquote{bandwidth of the fingers}. Physically, the HyFinBall device size follows Zhai et al.'s FingerBall, which had a single button activated by squeezing \cite{Zhai1995}. Our HyFinBall interface uses multiple buttons and is two-handed, following Shaw and Green \cite{Shaw1994}. Ulinski et al. \cite{Ulinski2007} use buttonballs similar to ours, but their system does not contain any of our HyFinBall hybrid UI concepts. 


Most input devices and corresponding interaction techniques that provide spatial input use either held devices or body-tracking, but not both, like the HyFinBall UI. There are some exceptions. For example, the touch mouse contains a multi-touch surface on the top of the mouse \cite{Bi2011}. However, to our knowledge, there has been relatively little development and experimentation with UIs that support 2D and 3D handheld devices while simultaneously enabling 2D/3D hand and finger-tracking. The goal of the HyFinBall with Finger-Tracking interface is to explore this design space. 

Section ~\ref{sec:fatigue} details HyFinBall's planar-3DOF and 6DOF auto-switching technique. Mapes and Moshel \cite{Mapes1995} use an HMD with 6DOF tracked pinch gloves and a physical surface. A pair of 3D cursors is positioned roughly corresponding to the position of the user's hands. When the hands rest on the surface, they are supported, and the pair of pinch-gloves essentially acts like a pair of 3-button mice. However, the display of the 3D cursors remains the same regardless of hand position. In contrast, in the HyFinBall planar-3DOF mode, if the user rests the buttonball on the desk, it changes both the cursor display and the interaction techniques available. This difference is motivated in part due to the display system difference, i.e,. HMD in Mapes and Moshel vs. desktop VR here. 

Hybrid user interface (HUI) refers to a UI with multiple methods for spatial input, frequently supporting both bimanual or unimanual interaction and 2D and 3D interaction. Benko et al. \cite{Benko2005} combine a multi-touch 2D surface with hand and finger 3D gestures and 3D interaction in an augmented reality system. They coined the terms HUI and cross-dimensional gestures.

Some earlier devices support a similar notion of cross-dimensional interaction. The Logitech 2D/6D Mouse and VideoMouse \cite{Hinckley1999} are a single device that supports both 6DOF mode and planar-3DOF mode. However, in neither system was this concept extensively developed into a hybrid 2D/3D UI, nor was bimanual interaction supported. No user studies similar to ours were performed. 

The utility of confining the motion of a 6DOF device to a physical plane, such as a held tablet, to reduce the physically manipulated DOFs has been demonstrated \cite{Bowman2008}. However, these prior works do not use a significant displacement between the physical device and its representative 2D or 3D cursor (as in \cite{Shaw1994}), and neither of these works' UIs implements the 6DOF to planar-3DOF mode switching found in the HyFinBall interface.

HyNet is an HUI system for desktop-based navigation \cite{Massink1999}. This work uses a mouse for 3D navigation with a conventional desktop system. However, the system only uses 2D GUIs with 2D UIs and does not provide a solution for 3D visualizations and VR systems. The authors also introduce a programming abstraction for the HUI, with traditional desktop-based systems that use conventional mouse and keyboard inputs. The HUI addresses both theoretical abstraction and 3D input modalities. 

HybridDesk \cite{Alencar2011} combines 2D and 3D interactions with a tracked Wiimote and WIMP interface for an oil platform visualization. There are three UIs in HybridDesk used to evaluate their HUI techniques: VR-Nav for navigation and selection, VR-Manip for manipulation, and the traditional WIMP UI. Magic Desk \cite{Bi2011} utilizes multi-touch input, a mouse, and a keyboard within a traditional desktop environment for unimanual and bimanual interactions. The authors explore suitable physical positions of multi-touch input relative to the user during the experiment. Althoff et al. present a multimodal interface for navigation in arbitrary virtual VRML worlds \cite{Althoff2001}, which uses a mouse, keyboard, joystick, and multi-touch input. However, their environment was limited to 2D visualizations and 2D interactions. The Slice WIM interface, which uses a multi-touch table with a head-tracked, stereoscopic wall screen display for a medical imaging volumetric dataset \cite{Coffey2011}, allows multi-touch interaction on the table to control 3D data using two widgets.

\begin{figure}[t]
\centering	
\includegraphics[width=0.9\columnwidth]{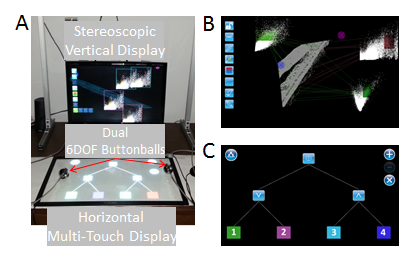}
\caption{HyFinBall UI: Fish-tank VR Setup (A) Screen capture of vertical display (B) and Horizontal Display (C).}
\label{fig:twohanded}
\end{figure}

\section{Hyfinball: a hybrid user interface}

Rich 3D visualizations involve coordinated views of both 2D and 3D components. The HyFinBall interface allows users to use 6DOF devices and 3D finger-tracking for 3D UI tasks while using planar-3DOF devices and multi-touch for 2D UI tasks--all without having to drop-and-pickup different physical devices. The HyFinBall system presents 3D graphics as well as tightly visually coupled 2D graphics on a vertical display, a head-tracked stereoscopic display. The system presents other 2D graphics, which are less tightly coupled to the 3D graphics, on a horizontal, multi-touch display. 

The HyFinBall system uses generic-shaped buttonball 6DOF devices that remain in the user's hands for relatively long durations to minimize device acquisition time penalties. This is opposed to using multiple, specially shaped 3D handheld devices that must be put down and picked up repeatedly. We suggest that for rich 3D visualization applications (as opposed to VR applications training users to use real-world tools), a pair of generic devices (or perhaps a few devices of different but generic shapes \cite{Moehring2011}) is preferable.

The HyFinBall system specifically uses small form-factor precision-grasped buttonballs. The generic sphere shape supports their usage in both 6DOF mode (Figure 1A) and planar-3DOF mode (Figure 1B). Their small size allows the index fingers and thumb to perform 3D finger gestures (Figure 1C) and multi-touch interaction (Figure 1D) while palming the buttonball.

The HyFinBall workstation uses Nvidia 3D Vision glasses and a vertical 24" 3D monitor, a Polhemus Fastrak for head and buttonball tracking, PQLab's 24" multi-touch frame with top-down projected horizontal display, and 3Gear software with dual PC Kinects for 3D finger tracking. Note, our using a top-down projector is a historical happenstance. We lacked a machine shop to custom-build a table with an inlaid LCD panel. Also, at the time \cite{Cho2013}, electromagnetic tracking was necessary for a small form-factor buttonball, and metal inside a horizontal LCD panel would preclude using an EM-tracked buttonball on the table surface.
Other hardware implementation combinations are possible. The results of this paper's user studies of the HyFinBall design and UI are independent of the hardware implementation.


Below, we briefly summarize a terrain analysis application, DIEM-VR, that uses and helps motivate the HyFinBall design and that is used in Experiment 1.

\subsection{DIEM-VR with 6DOF and planar-3DOF Modes}
DIEM-VR is a terrain analysis application. The vertical stereo display shows a patch of 3D terrain, specifically LIDAR scans from the North Carolina coast over a 10 year period. The right side of the screen displays various 2D icons. The user can create 2D scatter plots (appearing at zero stereo-parallax) that overlay 3D terrain. Each scatter plot point corresponds to a point in the terrain mesh. The scatter plots are a feature-space representation of the terrain. The user can choose various data for a plot's X-Y axes. For example, the Y-axis might be a plot point's terrain height, while the X-axis might be a plot point's average local gradient in the mesh. The user can create, destroy, and move the plots. The user brushes points in the plots, which automatically highlight the corresponding points in the terrain mesh using a color coding (see Figure 2B). A "line-net" can be displayed that draws lines from the selected 2D plot's points to the corresponding terrain mesh vertex. This further visually correlates the selection across the 2D plots and 3D terrain. Because this tightly visualizes the scatter plots to the 3D terrain, the plots are kept on the vertical display.

\subsection{Fatigue - Elbows-Resting vs. Hands-Resting }\label{sec:fatigue}
Shaw and Green \cite{Shaw1994} advocate adding a user-adjusted translation offset between the 6DOF button device and the 3D cursor in their two-handed desktop VR system. This allows the user to keep her elbows resting in her lap, or on the desk or chair arm, to combat the common arm fatigue problems in VR interfaces. We include this offset in all our Fish-tank VR systems. However, in prior experimental work \cite{Ulinski2007} and in our formative evaluation of the HyFinBall interface, we found that while keeping elbows resting on a surface reduces fatigue compared to the na\"{i}ve 'arm's outstretched' approach, even with the translation offset addition such 6DOF bi-manual input is still more fatiguing than using a regular 2D mouse. The reason is that with a mouse, the hand--not just the elbow--rests on a surface.

Therefore, we developed the HyFinBall UI with an "auto-mode switch" between 6DOF and planar-3DOF mode to allow the user to perform one or two-handed 3D interactions as well as one or two-handed 2D interactions with her hand(s) resting on the desk. When a buttonball is up, off the desk, 6DOF mode is engaged, and a 3D transparent cursor appears on the vertical display. When the buttonball is touching the desk, planar-3DOF mode is engaged, and a 2D transparent cursor appears at zero-parallax and is controlled by the buttonball like an absolute, position-controlled mouse. The left and right buttonballs switch independently between 6DOF and planar-3DOF modes. When in 6DOF mode, the 3D cursors interact only with 3D objects (terrain, 3D selection boxes). When in planar-3DOF mode, the 2D cursors interact only with 2D objects (scatter plots, menus, and icons). 

Depending on the modes being used, different eye separations are automatically used for stereo rendering. If only one HyFinBall is in planar-3DOF mode and is performing 2D interaction, then the modeled eye separation is cut in half. If both HyFinBalls are in planar-3DOF mode and performing 2D interactions, eye separation is set to zero. The eye separation changes are animated over the recommended 2s time period \cite{Ware2012}. 

Experiment 1 contains a 6DOF-only condition that uses image-plane interactions \cite{Bowman2004} for 2D interactions. Based on prior experience, we hypothesize that this condition will yield more arm fatigue and be slower than the HyFinBall interface when the user must perform both 2D and 3D tasks or just 2D tasks. Of course, there is a trade-off. In the HyFinBall interface, the user must place the buttonball on the table to switch from a 3D task to a 2D task, while in the 6DOF-only mode, a buttonball does not need to be placed on the desk.

\subsection{HyFinBall+Finger-Tracking}

The HyFinBall finger-tracking interface further leverages the small form factor and precision grip of the HyFinBall to allow the user's free fingers to interact with 3D finger tracking and 2D multi-touch while holding the buttonball.

We integrated the 3Gear 3D hand and finger tracking within the HyFinBall interface for 3D interactions with the 3D visuals and we integrated a PQLab's multi-touch frame with a projected horizontal display for user interaction with additional 2D GUIs (Figure 2A). At the time of the user study, the 3Gear tracking lacked adequate robustness for formal evaluation, so this paper only evaluates the HyFinBall integration of multi-touch. However, to the degree our results show that users can freely manipulate two fingers for multi-touch while holding the buttonball, we believe users would be equally capable of performing 3D finger gestures.

Conceptually, the HyFinBall UI uses the horizontal display for pure 2D interactions, whereas the vertical display is used for tightly visually coupled 2D and 3D visualization. We specifically chose to touch enable the horizontal display rather than the vertical one, to maintain a hands-resting posture during the multi-touch interaction. For example, DIEM-VR displays an interactive Boolean expression tree on the horizontal multi-touch display (Figure 2C). The tree describes a Boolean combination of data points built from the points brushed within multiple scatter plots on the vertical display.

This paper's Experiment 2 evaluates a user's ability to switch between 3D interactions with 6DOF buttonballs and 2D interactions with the multi-touch surface while palming the buttonball. Again, the core HyFinBall tenant is to allow the user to use the best spatial input technology for 2D input when performing 2D tasks and the best spatial input technology for 3D input when performing 3D tasks while avoiding device switching time. In Experiment 2, our hypothesis is that giving the user the ability to seamlessly flow between 3D interaction via buttonball and 2D interaction via multi-touch will lead to faster overall task completion for a combined 2D+3D task compared to a hardware interface that requires the user to physically switch devices.

\section{Experiment 1: 6DOF and planar-3DOF mode switch}

Experiment 1 uses a subset of interaction techniques within DIEM-VR. The three interaction technique conditions are:

\begin{description}
\item[C1:] \textbf{HyFinBall:}  Auto-switching between 6DOF and planar-3DOF provides 3D and 2D UIs.
\item[C2:] \textbf{6DOF-Only mode:}  3D UI is the same as in HyFinBall but 2D interactions done with image-plane techniques.
\item[C3:] \textbf{6DOF+Mouse:} 3D UI is the same as in HyFinBall but 2D interactions done with one mouse.
\end{description}

For the planar-3DOF sub-mode of C1, we calibrate planar-3DOF 2D cursor movement to require the same distance of physical motion as the mouse with mouse acceleration disabled in C3. In Experiment 1, DIEM-VR displays five menu icons, one patch of terrain, and a scatter plot on the vertical screen (Figure \ref{fig:twohanded}A and \ref{fig:twohanded}B). Experiment 1 does not use the multi-touch and 3D hand and finger interfaces.

3D tasks of Experiment 1 include: 

\begin{description}
\item[T1$_{3D}$:] 7DOF navigation (pose + scale)
\item[T2$_{3D}$:] Creation and manipulation of 3D selection boxes 
\end{description}

We refer to the mix of T1$_{3D}$ and T2$_{3D}$ as a ``3D Trial'', symbolized T$_{3D}$. Note, during a 3D Trial, the user freely switches between navigation and box manipulation.

Navigation uses a uni-manual scene-in-hand metaphor \cite{Ware1988} with a separate rate-controlled, cursor-centered scale \cite{Robinett1992}. Selection boxes are created using the bi-manual ``two-corners asynchronous'' technique of Ulinski et al \cite{Ulinski2007}. During a trial, the user is prompted with a red 3D selection box on the terrain. The user must both (1) navigate and (2) create and manipulate a 3D selection box to make a matching selection box. The created selection box corners must match corresponding target selection box corners with a tolerance of $\pm10$\% of the cursor box size. 

2D tasks of Experiment 1 are as follows:

\begin{description}
\item[T1$_{2D}$:] \textbf{menu icon selection:}

A red outlined rectangle appears on one of menu icons to prompt the user to select the icon. 
\item[T2$_{2D}$:] \textbf{scatter plot axis selection:}

A red outline appears on one of the scatter plot's data axis icons. The user selects this icon to change the variable plotted on that axis.
\item[T3$_{2D}$:] \textbf{relocating a scatter plot:}

A red outlined scatterplot appears in a random position, and then the user moves the target scatterplot to a position indicated by an objective rectangle. 

\item[T4$_{2D}$:] \textbf{brushing points a scatter plot:}

A red rectangle appears inside the target scatter plot, bounding a set of plot points. The user needs to draw a selection rectangle matching the red target. 
\end{description}

Each 2D Trial presents four 2D tasks, one of each of the above 4 types, in sequence. The order of 2D task presentation is randomized per 2D Trial.  For brevity, we denote a 2D Trial as T$_{2D}$.

\begin{figure*}[t]
\centering	
\includegraphics[width=0.8\textwidth]{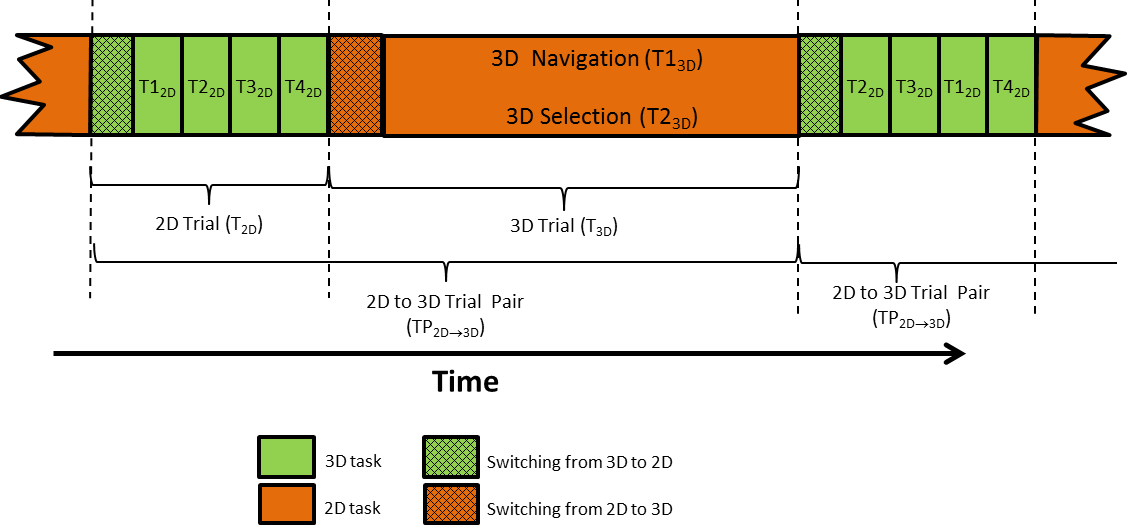}
\caption{Trials and Tasks}
\label{fig:trial_sequence1}
\end{figure*}

Trials are always generated in Trial Pairs. A Trial Pair contains one 2D Trial and one 3D Trial. Each Trial Pair can present the 2D Trial and 3D Trial in one of two possible orders: a 2D-to-3D Trial or a 3D-to-2D Trial. An equal number of each Trial Pair type is presented to each user. The choice between the two types is, however, random. Figure \ref{fig:trial_sequence1} illustrates a Trial Pair subdivided into a 2D Trail and a 3D Trail, which are each further subdivided as described above.  The cross-hatched time blocks are the duration when the user might have to switch between 2D and 3D input methods. In the HyFinBall condition, this represents placing a buttonball on the surface or lifting it off the surface. In the 6DOF+Mouse condition, this represents swapping the mouse for a buttonball. For the 6DOF-Only condition, there would be no 'switching' time since 2D and 3D UIs both use the buttonballs in an elbows-down-hands-up posture.

\begin{figure*}[t]
\centering	
\includegraphics[width=0.8\textwidth]{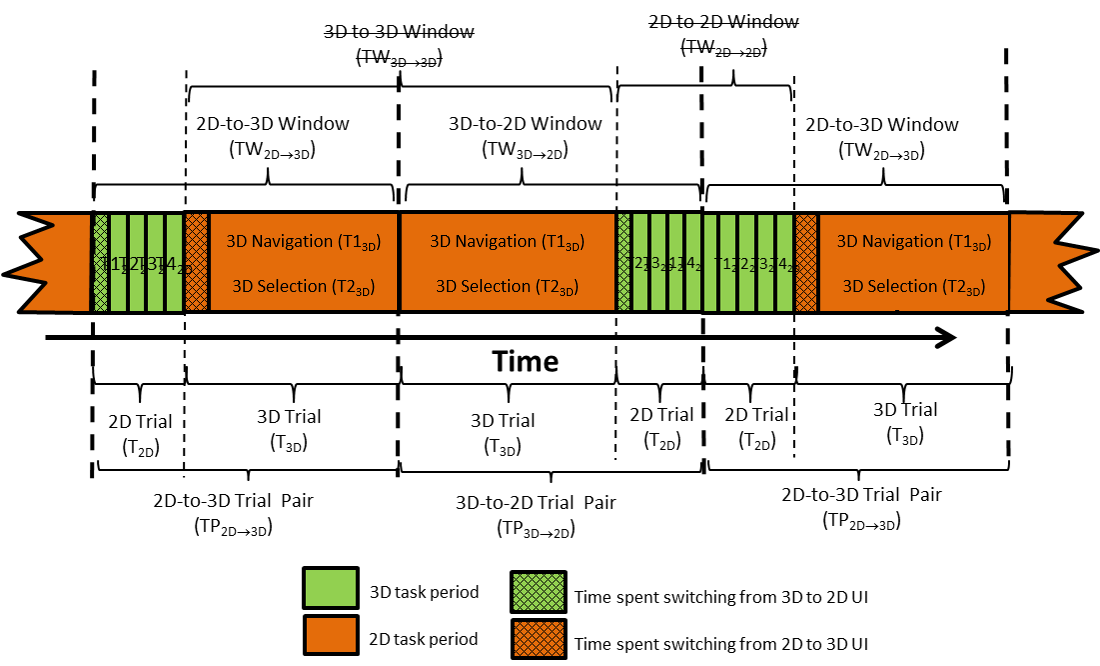}
\caption{Trials and Windows}
\label{fig:trial_sequence2}
\end{figure*}

In order to examine the effects of device-acquisition time, during analysis, we use a sliding window of two trials and examine each successive pair of trials. Depending on the particular sequence of Trial Pairs, each "Trial Window" can be of one of four types: 2D-to-3D window, 3D-to-2D window, 2D-to-2D window, or 3D-to-3D window. The total completion time for all 2D-to-3D windows per user is the 2D-to-3D completion time. The total completion time for all 3D-to-2D windows per user is the 3D-to-2D window completion time.  We calculate and examine these separately to compare the device switching penalty effects between the two cases.

Figure \ref{fig:trial_sequence2} illustrates the relationship between several Trial Pairs and several Trial Windows. The completion times for Trial Windows whose labeled text is in strike-through font require no 2D/3D UI switching. Again, only Trial Windows whose labeled text is in normal font (where device switching may occur) are accumulated into the analyzed 2D-to-3D and 3D-to-2D total times.


\subsection{Pilot Study}

%
%
In our pilot study for Experiment 1, we included 5 device conditions: 

\begin{description}
\item[C1$_P$:] HyFinBall with auto-switching between 2D and 3D 
\item[C2$_P$:] 6DOF-only mode
\item[C3$_P$:] 6DOF+mouse (dual 6DOF buttonballs with a mouse)
\item[C4$_P$:] mouse-only
\item[C5$_P$:] planar-3DOF-only
\end{description}

The 2D device conditions (mouse only and planar-3DOF buttonballs) proved significantly slower for 3D tasks during pilot tests by a factor of roughly 2 to 3.

For 3D interaction, the 2D device conditions, mouse-only and planar-3DOF-only, work as follows. For (7DOF \cite{Robinett1992}) navigation, the user can select a POI on the terrain. The mouse wheel scales up/down relative to the POI. In rotate mode, the camera rotates about the POI using Arcball \cite{Shoemake1992}. In translate-xy mode, mouse motion moves the camera parallel to the projection plane. In translate-z mode, mouse y-motion moves the camera perpendicular to the projection plane. For planar-3DOF buttonballs, the dominant buttonball mimics the mouse interface, but lacking a wheel, scaling is a separate mode controlled by y-motion. 
For 3D selection box creation, the user first selects two points on the terrain surface to define a square. Two additional vertical mouse motions extrude the box above and below the initial plane. Further box adjustments use a combination of Arcball object rotation and 3D manipulation handles \cite{Bowman2004} on the box for additional translation and scaling. In the planar-3DOF condition, the non-dominant buttonball performs this 2D mouse interface.

We ran 2 participants (non-authors) in a within-subjects experiment across all 5 device conditions. We found that the 2D device conditions (mouse, planar-3DOF) perform very poorly relative to the 3D devices (6DOF only, 6DOF+Mouse, HyFinBall) during the 3D tasks.  The 3D task completion times for the mouse ($\mu=65.89$) and planar-3DOF buttonballs ($\mu=100.88$) were significantly slower than for the 3D devices ($\mu=34.94$) by factors of 2 and 3. Trials were in blocks by device. 2D device blocks were interleaved with the 3D device blocks, so the poor 2D device performance was not due merely to fatigue. This is similar to previous studies \cite{Schultheis2012} that use more complex, multi-step 3D interactions and find relatively poor performance of 2D devices. From our anecdotal observation, the main culprit of the performance difference was the 3D box selection rather than navigation. The experiment took slightly over 2 hours per participant, and they complained of visual and physical fatigue. Further, they complained heavily regarding having to perform the 3D selection box task with either a 2D device condition after they experienced the 3D bi-manual interface of conditions C1$_P$, C2$_P$, and C3$_P$. Based on this experience, we removed the mouse only and planar-3DOF conditions from Experiment 1. Experiment 1 C1 through C3 were the same as pilot study conditions C1$_P$ through C3$_P$.

\subsection{Participants and Design}
Twelve participants performed 30 trials each (10 trials (10 $\times$ 3D task and 10 $\times$ 2D task) $\times$ 3 UI conditions) in a within-subject design over HyFinBall UI, 6DOF+mouse, and 6DOF only. Eleven participants are from the Computer Science department and one from Public Health Sciences; 7 are males, and 5 are females. All participants have (corrected) 20/20 or higher vision, no disability using their hands and fingers, and passed a stereopsis test. All participants have high daily computer usage (6.25 out of 7). Eleven participants have experience with a 6DOF buttonball UI from a previous 1-hour navigation user study. The three UI conditions were presented in counter-balanced order across all participants.


%

We evaluated four completion times corresponding to 3D Trials and 2D Trials (Fig. \ref{fig:trial_sequence1}), and 3D-to-2D Window Trials and 2D-to-3D Window Trials (Fig. \ref{fig:trial_sequence2}).



Our primary hypotheses for Experiment 1 are:
\begin{enumerate}[label=\textbf{H\arabic*:},ref=H\arabic*]
\item \label{itm:H1} HyFinBall is expected to have faster overall completion time than other UIs. This is due to auto-switching (no device acquisition time compared to 6DOF+Mouse) and providing a better mapping for 2D interaction (planar-3DOF submode) than the image-plane techniques required for 6DOF-only.
\item  \label{itm:H2} HyFinBall (\textbf{H2-A}) and 6DOF+Mouse (\textbf{H2-B}) are expected to have faster completion time than 6DOF-only for 2D tasks. This is due to that they both have a more natural 2D interaction sub-mode than 6DOF-only image-plane techniques.
\item   \label{itm:H3} UI condition is expected to have no significant effect on completion time for the 3D task. 

Note, it is possible that the cognitive load of the novelty of the HyFinBall condition -- in particular, the 6DOF and planar 3DOF mode switching -- would detrimentally hurt performance overall, including 3D tasks. Hence, despite the fact the 3D task UI's are the same across all conditions, the HyFinBall condition could have a detrimental effect overall. However, our experience as expert users and designers of the HyFinBall UI suggests HyFinBall's novel aspect would not detrimentally affect performance. 

\item    \label{itm:H4} HyFinBall is expected to have faster completion time for cross-dimensional trials (3D-to-2D and 2D-to-3D) than other 2 UIs.  

This is due to HyFinBall's the lack of device acquisition time compared 6DOF+Mouse and its use of 2D input devices for 2D interaction compared to the image-plane techniques of 6DOF-only.
\end{enumerate}

\subsection{Results}

We primarily use repeated measures ANOVA (Analysis of Variance). The F tests that are reported use $\alpha=.05$ for significance, and we use the Geisser-Greenhouse correction to protect against possible violation of the assumption of sphericity. The post-hoc tests are least significant differences (LSD) tests with $\alpha=.05$ level for significance.

\subsubsection{Quantitative}

There is no interaction effect of UI $\times$ TrialType (T$_{2D}$ vs T$_{3D}$) on completion time (3 $\times$ 2 repeated measures ANOVA (UI $\times$ TrialType), $p=.165$).

Results confirm hypothesis \ref{itm:H1}.  There is a main effect on completion time of UI condition ($F(2,22)=7.857, p=.008, \eta_{p}^{2}=.417$). LSD tests show completion time of HyFinBall ($\mu=27.72, \sigma=8.66$) is faster than 6DOF+Mouse ($\mu=31.08, \sigma=11.03, p=.021$) and 6DOF-only ($\mu=34.57, \sigma=12.82, p=.010$). 

Regarding hypothesis \ref{itm:H1}, one-way repeated measures ANOVA for the 2D Trials shows a simple effect on completion time of UI condition ($F(1.359,14.944)=11.839, p=.002, \eta_{p}^{2}=.518$). LSD tests show completion time of 6DOF+Mouse is faster than 6DOF-only for 2D Trials ($p=.005$), confirming H2-A. HyFinBall is also faster than 6DOF-only ($p=.003$), confirming H2-B.  HyFinBall and 6DOF+Mouse were not significantly different. This indicates 2D Trial performance of HyFinBall is also faster than 6DOF-only ($p=.003$), confirming H2-B. HyFinBall and 6DOF+Mouse were not significantly different. This indicates 2D Trial performance of HyFinBall's planar-3DOF sub-mode and 6DOF+Mouse's mouse sub-mode is better than the 6DOF-Only image-plane modes. This also indicates 2D Trials performance of planar-3DOF is no worse than the mouse input. We observed in the 6DOF-Only condition, most participants had difficulty holding the buttonball in a fixed position in the middle of the air when they were performing image-plane 2D interaction. This was despite the fact that they could rest their elbows due to the translation cursor offset.

Regarding \ref{itm:H3}, there is no simple effect of UI on completion time for 3D Trials ($p=.088$). This is consistent with \ref{itm:H3}. It is, of course, statistically possible that a larger study would find a statistically significant difference. But importantly, it suggests that HyFinBall's novel auto-mode switch does not detrimentally affect 3D performance due to additional cognitive load.

The 3 $\times$ 2 repeated measures ANOVA for cross-dimensional tasks (3D-to-2D Trial Windows and 2D-to-3D Trial Windows) shows a main effect on completion time of UI condition ($F(2,22)=7.606, p=.003, \eta_{p}^{2}=.409$). Completion time of HyFinBall ($\mu=55.31, \sigma=8.82$) is faster than both 6DOF+Mouse ($\mu=63.96, \sigma=12.5, p=.010$) and 6DOF-only ($\mu=69.02, \sigma=18.17, p=.007$). This confirms hypothesis H4. Also, 6DOF-only and 6DOF+Mouse do not differ ($p=.155$).

\begin{table}[t]
 \caption{Average($\mu_{CT}$) and standard deviation ($\sigma$) of task completion time of different UIs.}
 \centering
 \begin{tabular}{c}
 \includegraphics[width=0.9\columnwidth]{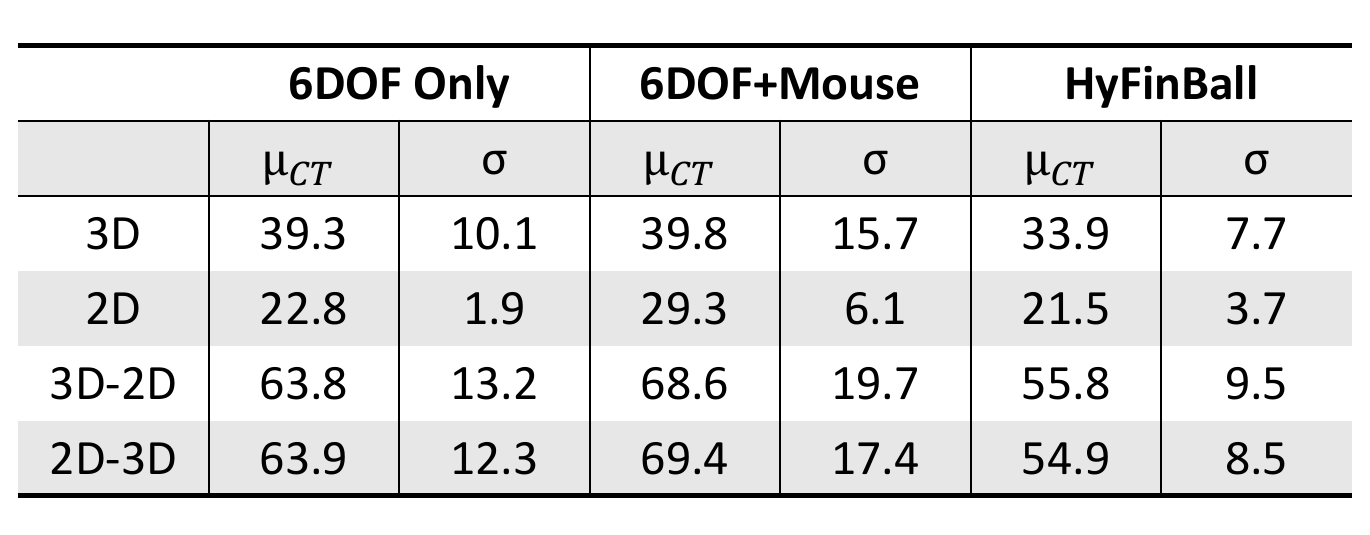}
 \end{tabular}
 \label{tab:table2}
\end{table}

\begin{figure}[t]
\centering	
\includegraphics[width=1.0\columnwidth]{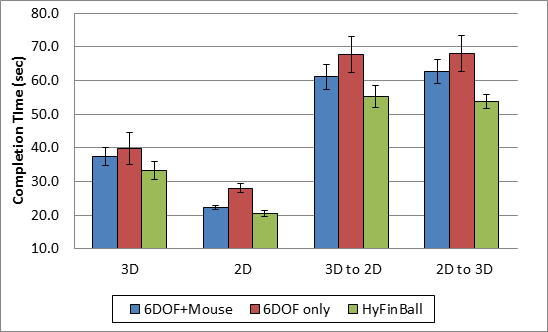}
\caption{Task completion time of  UI condition for tasks. Error bars indicate $\pm$1.0 standard error}
\label{fig:graph1}
\end{figure}

The one-way repeated measures ANOVA for the 3D-to-2D task shows a main effect on completion time of UI condition ($F(1.376,15.131)=6.571, p=.015, \eta_{p}^{2}=.374$). Completion time of HyFinBall is faster than both 6DOF+Mouse ($p=.010$) and 6DOF-only ($p=.018$). However, there is no difference in completion time between 6DOF-only and 6DOF+Mouse ($p=.169$). 

The one-way repeated measures ANOVA for the 2D-to-3D task show a main effect on completion time ($F(2,22)=7.359, p=.004, \eta_{p}^{2}=.401$). Completion time of HyFinBall is faster than both 6DOF+Mouse ($p=.013$) and 6DOF-only ($p=.008$). Same as the result of the 3D-to-2D task, there is no difference between 6DOF+Mouse and 6DOF-only ($p=.172$). The HyFinBall UI has the best task performance for cross-dimensional tasks. This supports hypothesis H4 that the overhead of switching HyFinBall sub-modes between 3D and 2D takes less time than changing physical input devices in the 6DOF+Mouse UI. Even though the 2D task performance of the mouse input is better than 6DOF-only, the device acquisition time penalty may reduce the overall performance of 6DOF+Mouse for cross-dimension tasks. 

\subsubsection{2D Tasks}

There is a main effect on 2D task completion time of 2D task types, T1$_{2D}$ through T4$_{2D}$ ($F(1.352,14.873)=69.241, p<.001, \eta_{p}^{2}=.863$). Completion time of T1$_{2D}$ Selecting Menu is faster than other 2D tasks ($p<.001$). Selecting Scatter-plot Axis Icons (T2$_{2D}$, SA) is faster than Relocating Scatter-plot (T3$_{2D}$, RS) ($p=.007$) and Brushing Points (T4$_{2D}$, BP) ($p<.001$). RS is faster than BP ($p<.001$). BP tends to be the slowest because it requires more accurate control and specifying 2 points.

\begin{figure}[t]
\centering	
\includegraphics[width=0.9\columnwidth]{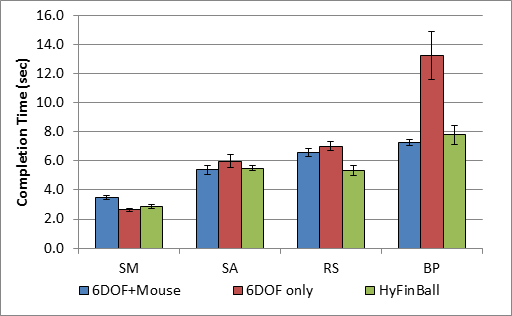}
\caption{Task completion time of UI condition for 2D tasks. Error bars indicate $\pm$1.0 standard error}
\label{fig:graph2}
\end{figure}

More subtly, there is an interaction effect between 2D task type and UI condition on completion time. The 3 $\times$ 4 repeated measures ANOVA (UI $\times$ 2D Task Type) is significant ($F(1.533,16.867)=9.773, p=.003, \eta_{p}^{2}=.470$). This indicates UI condition has different effects on completion time depending on the particular 2D task 
(see Figure \ref{fig:graph2} and Table \ref{tab:table2}). We had not anticipated this interaction. Hence, some 2D tasks' simple effects support H2 (a main effect hypothesis), and some do not, as discussed below.

\begin{table}[t]
 \caption{Average($\mu_{CT}$) and standard deviation ($\sigma$) of task completion time of different UIs.}
 \centering
 \begin{tabular}{c}
 \includegraphics[width=0.9\columnwidth]{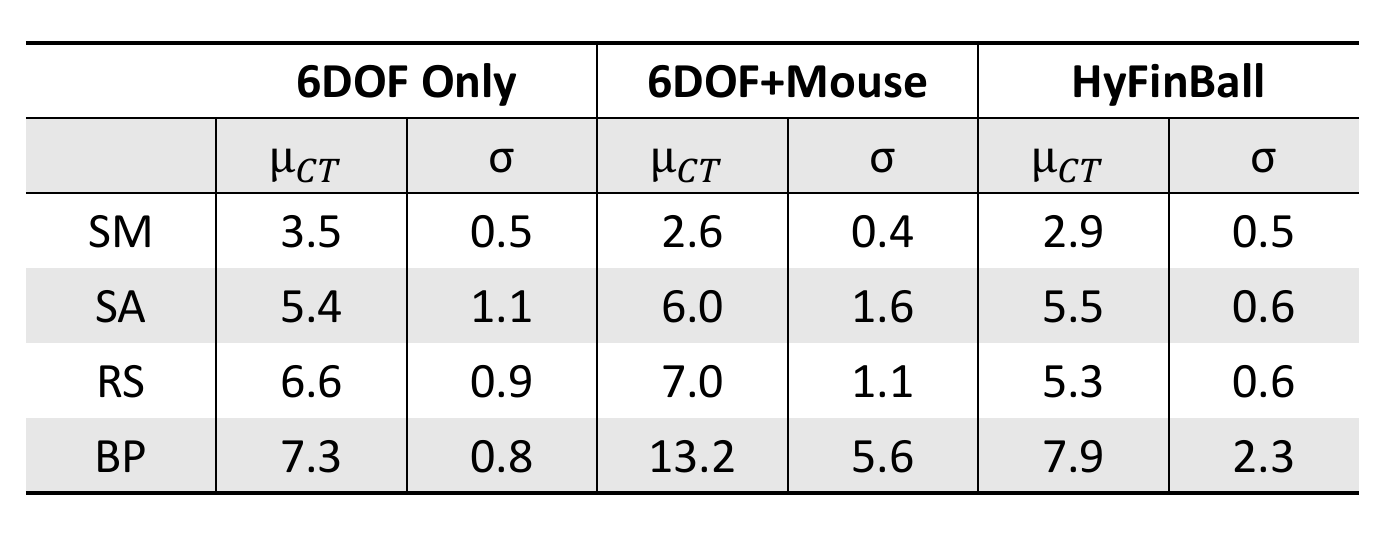}
 \end{tabular}
 \label{tab:table2}
\end{table} 

There is no simple main effect for Selecting Axis (SA) across UIs.

There is a simple main effect on completion time of UI condition for the Selecting Menu (SM) task ($F(2,22)=9.618, p=.001, \eta_{p}^{2}=.466$). Completion time of both HyFinBall's planar-3DOF sub-mode ($p=.011$) and 6DOF-only's image-plane UI ($p=.001$) is faster the the 6DOF+Mouse sub-mode. There is no difference between 6DOF-only and HyFinBall ($p=.289$). The first item indicates the benefit of using dual 2D cursors in HyFinBall's planar-3DOF sub-mode versus 6DOF+Mouse's single 2D cursor. This is in line with research comparing dual-mouse UIs to single-mouse UIs. The fact that the mouse performs slower than the 6DOF-only is likely also due to the fact that 6DOF-only allows either hand to perform 2D operations. The 6DOF-only left-hand 3D cursor likely remained close to the left side of the screen and the menu. It appears that the 6DOF-only dual-handedness offset the possible deficit of having to use image-plane interaction when competing with the 6DOF+Mouse mouse sub-mode.

There is a simple main effect on completion time in the Relocating Scatter-plot (RS) task ($F(2,22)=10.917, p=.001, \eta_{p}^{2}=.498$). Completion time of HyFinBall's planar-3DOF sub-mode is faster than both 6DOF-Only ($p=.003$) and 6DOF+Mouse ($p=.003$). Completion time of the mouse and 6DOF-only do not differ ($p=.223$). HyFinBall's planar-3DOF sub-mode outperforming 6DOF-only's image-plane techniques was expected in H2. However, H2 expected 6DOF+Mouse mouse sub-mode to outperform 6DOF-only for 2D tasks (the mouse being more natural for 2D than 6DOF-only's image-plane techniques). The opposite result occurred, likely because the 6DOF-only benefited over the mouse from its dual 3D cursors more than it lost from its image-plane methods. (This also occurred with the Select Menu sub-task (SM)). Again, either 3D cursor could be used to relocate the target scatter plot, and by the user choosing the closer 3D cursor closest to the target plot, she can initiate the RS task sooner than with a single mouse cursor.

Compare the Brushing Points and Relocating Scatter-plot results to those of the Selecting Menu (SM) task. In both BP and RS, the best performance came from a 2D device, but in BP, the mouse dominated, while for RS, the planar-3DOF dominated. It is unclear why the mouse and planar-3DOF results are reversed. The planar-3DOF condition has a potential 2D advantage of having two cursors. Because of the screen layout, with the planar-3DOF buttonballs, users tend to use the left cursor for Select Menu and the right cursor for Relocate Scatter-plot and Scatter-plot Axis. Having dual cursors might reduce total cursor travel compared to having a single cursor because with a single cursor, the user must traverse the cursor across the left and right sides of the screen. But if this were the only explanation for the difference in performance, one would not expect the mouse to have dominated in the Select Menu task.

Finally and most noticeably, there is a simple main effect on completion time of UI condition for the Brush Points (BP) task, ($F(1.257,13.829)=10.437, p=.004, \eta_{p}^{2}=.487$). Completion time of 6DOF-only is slower than the 6DOF+Mouse mouse sub-mode ($p=.009$) and the planar-3DOF ($p=.005$). However, the completion time of 6DOF+Mouse mouse sub-mode and HyFinBall planar-3DOF sub-mode do not differ ($p=.453$). From our SM and RS sub-task results, we might have expected HyFinBall's planar-3DOF sub-mode to outperform 6DOF+Mouse's mouse sub-mode due to its dual 2D cursor advantage. However, dual cursors generally only help reduce the start time of a 2D task (with two cursors, it is more likely that one cursor is closer to the target than with one cursor). Once the task is started (at least in all our 2D sub-tasks), only one 2D cursor is used. BP generally took longer overall than the other 3 sub-tasks. Hence, the improvement in task start time due to dual cursors is likely a smaller percentage contribution to overall time reduction. It is also possible that since the relative DPI of the naive mouse is higher than the calibrated HyFinBall's planar-3DOF mode 2D cursors, the naive mouse's precision advantage comes to the fore in the BP condition.

\subsubsection{Qualitative}

This section returns to the overall results, focusing on subjective ones. On a 7-point Likert scale, user rating of arm fatigue (\textit{1 no fatigue} through \textit{7 very painful}) is significantly different across UI conditions. The one-way repeated measures ANOVA shows a main effect on arm fatigue of UI condition ($F(1.289,14.176)=10.333, p=.004, \eta_{p}^{2}=.484$). Participants felt more arm fatigue with 6DOF-only ($\mu=3.5, \sigma=1.45$) than 6DOF+Mouse ($\mu=2.25, \sigma=0.97, p=.033$) and HyFinBall ($\mu=2.0, \sigma=1.04, p=.013$). However, arm fatigue rating of 6DOF+Mouse and HyFinBall do not differ significantly ($p=.537$). The plausible explanation is that during 2D tasks, while users' elbows generally rest in all conditions, their hands only rest in HyFinBall and 6DOF+Mouse conditions, and not the 6DOF-only condition.  

It is precisely the goal of the HyFinBall design to achieve a reduced fatigue benefit similar to 6DOF+Mouse over 6DOF-only, while avoiding the device acquisition time of 6DOF+Mouse in 2D/3D tasks (shown in the previous section).

On a 7-point Likert scale, user rating of how accurately they can perform the 2D task (\textit{1: no accuracy} through \textit{7: very accurate}) is significantly different across UI conditions ($F(2,22)=4.602, p=.021, \eta_{p}^{2}=.295$). Users were more confident to perform 2D tasks accurately with HyFinBall ($\mu=6.0, \sigma=1.13$) than 6DOF-only ($\mu=4.58, \sigma=1.08, p=.028$). Subjective confidence rating of 6DOF+Mouse ($\mu=5.5, \sigma=1.38$) is slightly higher than 6DOF-only UI but not significantly so ($p=.307$). On the same 7-point Likert scale, user ratings of how accurately they could perform the 2D-to-3D and 3D-to-2D tasks do not differ significantly across UI conditions.

Overall, 10 out of 12 users answered that they thought HyFinBall is the easiest UI for the combined 2D/3D tasks, 2 answered 6DOF+Mouse. For the 2D task, 8 users answered that the mouse is the easiest, and 4 answered planar-3DOF. For 2D tasks followed by 3D tasks, 8 participants answered that HyFinBall is the easiest, 4 answered 6DOF-only. For the 3D task followed by the 2D task, 9 answered that HyFinBall is the easiest UI, and 3 answered 6DOF-only. 

When asked how much they prefer planar-3DOF buttonball or mouse for 2D tasks (\textit{-2: strongly prefer mouse, -1: somewhat prefer mouse, 0: neutral, +1: somewhat prefer planar constraint buttonball, +2: strongly prefer buttonball}), 4 users answered they strongly preferred planar-3DOF buttonballs, 5 answered they somewhat preferred planar-3DOF buttonballs, 1 answered strongly preferred a mouse, and 2 answered they somewhat preferred a mouse.

\section{Experiment 2: Multi-Touch While Palming}
\begin{figure}[!t]
\centering	
\includegraphics[width=0.9\columnwidth]{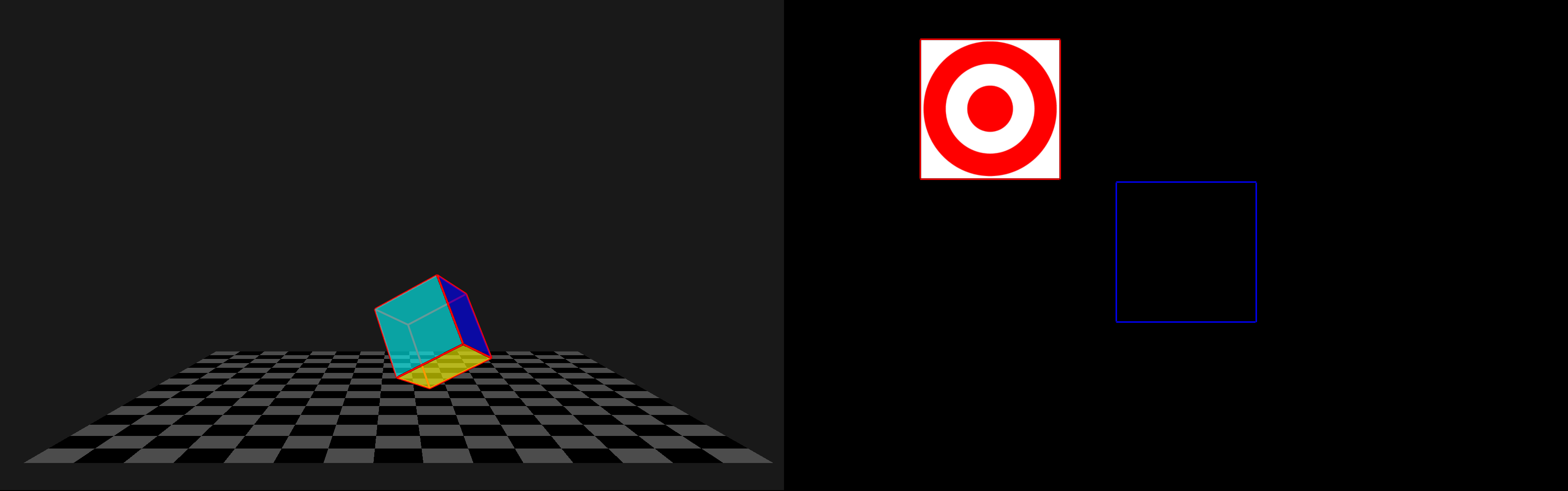}
\caption{3D selection task on the vertical screen (left) and multi-touch task on the horizontal screen (right).}
\label{fig:touch}

\end{figure}

The HyFinBall UI aims to allow the user to seamlessly switch between 4 modes: 6DOF buttonball (3D), planar-3DOF buttonball (2D), multi-touch (2D), and hand and finger tracking (3D) -- all without a device acquisition time penalty. In prior work that presents the DIEM-VR HyFinBall system, we describe and informally demonstrate with a video an expert user using and switching between all 4 modes during a terrain analysis session \cite{Cho2013}. However, the recorded users were all expert users. Experiment 1 formally evaluates the 6DOF to planar-3DOF mode switching. Next, Experiment 2 evaluates the usability of switching between 6DOF buttonball input and 2D multi-touch while ``palming'' the buttonball. This avoids dropping and picking up the 6DOF device.  Palming leaves two fingers free (see Figure 1D) per hand. The goal of Experiment 2 is to demonstrate and formally test the viability of the palming technique and to compare it to an alternative user strategy where the user drops and picks up the 3D input device when switching between 3D and multi-touch input.

Recall, HyFinBall UI's multi-touch occurs on the horizontal display only (to maintain a hand resting posture) for interacting with 2D visuals. Generally, these visuals are only loosely coupled with the 3D data on the vertical display. In the DIEM-VR application (Section 3.1), multi-touch is used for manipulating a Boolean tree defining a feature-space combination of scatter plot selections. Experiment 2, however, uses a separate stand-alone application. We did this because the multi-touch surface in DIEM-VR is only used for one-finger touch operations. We desire to create a user task for Experiment 2 that requires dual-finger touch operations. Rather than artificially inserting a new dual-finger operation into the DIEM-VR application, we create a separate application for Experiment 2. Similar to Experiment 1, the user performs a series of tasks that require switching between 3D interaction and 2D multi-touch. The same hardware setup is used in Experiment 2 as in Experiment 1.

There are two device conditions. They differ in the user's required strategy only -- the actual input devices used are held constant:

\begin{description}
\item[M$+$B:] multi-touch while palming the buttonball.
\item[M$-$B:] multi-touch while not holding the buttonball. In the M$-$B condition, the user is instructed to put down and pick up the buttonball when switching between the 3D and multi-touch tasks.
\end{description}

While the HyFinBall design is bi-manual and each hand can independently switch between all 4 modes, Experiment 2 only uses one hand, the user's dominant hand, for all tasks. The two tasks are a 3D selection task on the vertical stereo screen and a multi-touch task on the horizontal screen (see Figure \ref{fig:touch}). For the 3D selection task, the user needs to use the 3D cursor of the dominant hand to select a colored cube. When the cursor is inside the cube, the color of the cube's outline changes from red to green. Then the user clicks a button to finish the 3D task. The box changes its position randomly for each 3D task.

Next, a target 2D box and a blue-wired 2D box appear on the horizontal multi-touch screen, and the vertical display goes blank. The first multi-touch task is translating a 2D target box to the center of the screen with a single finger (typically index finger). After finishing the translation task, the blue-wired box changes its orientation ($\leq \pm 45^{\circ}$) and scale (50\% and 200\% of the target box size) randomly. The second multi-touch task is changing the target box to the blue-wired box's orientation and size. The user uses two fingers (typically the thumb and index finger) for rotation and scaling.

Each trial requires the user to perform the 3D task followed by the 2D multi-touch task. There is a space on the desk below the horizontal multi-touch frame (7.5") to rest one's elbow. The user is instructed to drop the buttonball in an area outside the multi-touch screen in condition M-B. This is because the PQLab's frame would recognize a buttonball placed inside the frame as a touch input. In the full HyFinBall UI, the system ignores touch input if the buttonball rests in the rectangular extent of the PQLabs frame (as detected by the 3D tracked position).

We record task completion time for both 3D and multi-touch tasks, and 3D and touch response times. 3D response time is the time from the completion of the multi-touch task to the time the buttonball is 2" above the last position of the buttonball after the multi-touch task. Multi-touch response time is the time from the completion of the 3D task to the first touch. We also separate the multi-touch sub-task completion time to separate the translation and rotation phases.

Our primary hypotheses of Experiment 2 are:

\begin{description}
\item[H1:] Overall, M+B is expected to have a faster response time than M-B. 

Note, implicitly, this demonstrates our participant population can learn and use the palming technique with minimal training.
\item[H2:] M+B is expected to have faster task completion time than M-B for both 3D selection and multi-touch tasks.
\end{description}

\subsection{Participants}

Sixteen users performed 80 trials each (40 trials $\times$ 2 UI conditions) in a within-subject comparison of two UI conditions, palming and non-palming, for the combination 3D plus multi-touch task. Nine participants are from the Computer Science department, and 7 are non-CS, such as Education or English majors. Eight are male,s and 8 are females. All participants have (corrected) 20/20 or higher vision and no disability using their hands and fingers. All participants have high daily computer usage (6.64 out of 7 on a Likert scale). All participants have experience with multi-touch devices such as Apple iPad and Samsung Galaxy S (daily usage: 6.19 out of 7), and 10 of 16 participants have 3D UI experience. The average hand size of females is 2.86" $\times$ 6.53" (width $\times$ height), and the average of males is 3.47" $\times$ 7.22". 
The total average of sixteen users is 3.16" $\times$ 6.88". The two UI conditions were presented in counter-balanced order across all participants.

\subsection{Quantitative Results}
We use one-way repeated measures ANOVA for data analysis. The F tests that are reported use $\alpha=.05$ for the level of significance. 

\textbf{Response Time:} The result shows a main effect of UI condition on 3D response time ($F(1,15)=198.301, p<.001, \eta_{p}^{2}=.930$). 3D response time of M$+$B ($\mu=0.28, \sigma=0.08$) is faster than M$-$B ($\mu=1.02, \sigma=0.19$). This reflects device acquisition time for M$-$B.

There is a main effect of UI condition on multi-touch response time ($F(1,15)=45.283, p<.001, \eta_{p}^{2}=.751$). Touch response time of M$+$B ($\mu=0.95, \sigma=0.13$) is faster than M-B ($\mu=1.49, \sigma=0.35$). This reflects device drop time. As expected, M$+$B has a faster response time for both 3D and touch (hypothesis H1). This is best explained by the device acquisition time required in the M$-$B condition.

There is no difference between groups (female vs. male), which means hand size was not significant with our limited sample.

\textbf{Task Completion Time:} The result shows a main effect of UI condition on 3D task time ($F(1,15)=65.912, p<.001, \eta_{p}^{2}=.815$). As expected (hypothesis H2), 3D selection task completion time of M$+$B ($\mu=2.01, \sigma=0.47$) is faster than M$-$B ($\mu=2.84, \sigma=0.54$). This appears to be due to the device acquisition time in the M-B condition.

There is no main effect of UI condition on multi-touch task time ($p=.178$). Multi-touch task completion time are M$+$B ($\mu=6.04, \sigma=0.84$) and M$-$B ($\mu=6.27, \sigma=0.83$) does not differ. 
There is a main effect on task completion time for the single-touch task of UI condition ($F(1,15)=4.809, p=.044, \eta_{p}^{2}=.243$). Completion time of M$-$B ($\mu=1.58, \sigma=0.34$) is significantly faster than M$+$B ($\mu=1.68, \sigma=0.39$). This indicates that palming the buttonball slightly slows down touch interaction, even if only one finger is needed. 

There is no main effect on task completion time of UI condition for the dual touch, rotate\&scale task ($p=.089$). Task completion time are M$+$B ($\mu=3.38, \sigma=0.45$) and M$-$B ($\mu=3.21, \sigma=0.33$).

\subsection{Qualitative Results}
Participants took a post-survey questionnaire regarding subjective preferences. 10 users answered they prefer M$+$B for the overall trial (3D+multi-touch task), 5 preferred M$-$B, and 1 answered no preference. For the 3D selection task, 14 preferred the M$+$B condition, 1 answered M$-$B, and 1 answered no preference. Not surprisingly, for the multi-touch task alone, 16 answered they prefer M$-$B, and 5 answered M$+$B.

\section{Discussion}
Experiment 1 shows that the HyFinBall UI performs faster over the 6DOF+mouse UI for cross-dimensional tasks. The 6DOF+Mouse condition has the inherent deficit of device switching. Also, for 2D tasks, the mouse is only uni-manual while the planar-3DOF buttonballs are bi-manual. However, our 2D tasks did not explicitly take advantage of bi-manual 2D interactions beyond the fact that having 2D cursors may minimize cursor travel distance. On the other hand, the mouse resolution (DPI) is higher than the planar-3DOF buttonballs, and the mouse form-factor is more familiar. While these other variables probably play some role in the HyFinBall UI vs 6DOF+Mouse performance difference, the results clearly demonstrate that users can easily learn and use the HyFinBall concept of the 6DOF to planar-3DOF mode switch, and subjectively, 10 of 12 prefer it for cross-dimensional tasks. Further, the results are consistent with the hypothesis that removal of the device acquisition time in the HyFinBall UI can reduce task completion time in cross-dimensional tasks.

Experiment 1 also shows that the HyFinBall UI performs faster compared to the 6DOF+Only UI for cross-dimensional tasks. The results from the cross-dimensional task analysis and as well as the 2D task analysis, indicate the 6DOF-Only condition is slower because the image-plane techniques are slower than the planar-3DOF techniques for 2D operations. This is not surprising given that a standard mouse has generally been shown to perform better for 2D interactions due to arm stability, etc. Subjectively, for the combined task, no user preferred 6DOF-only to the HyFinBall.

We also evaluated how many times the user presses wrong buttons for 2D interactions across the UI conditions in Experiment 1, but did not report the significance in this paper. In short, the user presses the wrong buttons more frequently with HyFinBall than with the mouse for 2D tasks. Most likely, this could be due to the different form factor of the buttonball vs. the mouse. This error might be reduced by adding tactilely distinguishable shapes to the three buttonball buttons. 

Experiment 2 demonstrated no overall detrimental effect when the user is palming the buttonball on the multi-touch sub-task. Further, it shows better performance for the 3D task when palming the buttonball because of the lack of the device acquisition time penalty (0.28s vs. 1.02s). 

The sequence or duty-cycle of 2D vs. 3D tasks is important for the relative performance of HyFinBall's 6DOF/planar-3DOF auto-switching and HyFinBall's multi-touch while palming. If the user needs to do many 2D tasks and rarely 3D tasks, then the 6DOF+Mouse UI might do better than the HyFinBall due to the planar-3DOF's unaccustomed shape.  (The lower DPI might also have an effect, but this is an implementation artifact.) Undoubtedly, the best shape for a mouse and the best shape for a pure 6DOF device are not the same. There may be a best compromise shape that retains the 6DOF precision grip advantage and still allows the HyFinBall "finger-tracking while palming the device" trick. Note that the HyFinBall UI does not rule out a user using a mouse during periods of mostly 2D work and then switching to the HyFinBall UI during task periods that switch frequently between 2D and 3D.

\section{Conclusion and Future Work}

Complex 3D visualization applications often include integrated and coordinated 3D and 2D visualizations. The HyFinBall UI design aims to allow users of such applications to use 6DOF devices and 3D finger-tracking for 3D tasks while using mouse-like planar-3DOF input and multi-touch for 2D tasks--all without having to drop-and-pickup different physical devices. The HyFinBall UI is designed for a dual-screen Fish-tank VR setup and is designed to provide maximal physical support of the user's hands and arms during these cross-dimensional tasks.

This paper formally evaluated two key HyFinBall UI components: 6DOF/planar-3DOF auto-switch and buttonball palming with multi-touch. Users successfully learned the auto-switch feature, and when performing combinations of 2D and 3D tasks, they performed faster overall with the HyFinBall auto-switch than with other UI methods. Users also successfully learned to use the multi-touch while buttonball palming technique, and when performing a combination task using 2D multi-touch and 3D positioning, they showed faster performance when using the palming technique. 

As commercial 3D finger-tracking achieves the robustness of multi-touch panels, we will develop and evaluate buttonball palming techniques with 3D finger tracking. (We demonstrate this capability in \cite{Cho2013}, but limited tracking robustness had interfered with our pilot studies).  Exploring the range of other device form factors that also support the HyFinBall hybrid techniques is also an interesting avenue. A broader issue is exploring how the duty-cycle between 2D and 3D interaction periods affects user performance, as well as subjective preference for using devices designed for hybrid spatial input compared to using multiple devices, each of which is designed only for 2D or only for 3D.

\section*{Acknowledgments}
This work was supported in part by grant W911NF0910241 (PN 55836MA) from the U.S. Army Research Office.
%
%
%
%
%

\bibliographystyle{acm-sigchi}
\bibliography{hfb}




\end{document}